\documentstyle[aps,prl]{revtex}
\begin{document}
%\draft
\twocolumn[\hsize\textwidth\columnwidth\hsize\csname 
@twocolumnfalse\endcsname
\title{Dynamic vortex mass in clean Fermi superfluids and 
superconductors}
\author{N. B. Kopnin$^{(1,2)}$ and V. M. Vinokur$^{(2)}$}
\address{$^{(1)}$ L. D. Landau Institute for Theoretical Physics, 
117940
Moscow, Russia\\
$^{(2)}$ Argonne National Laboratory, Argonne, Illinois 60439}
\date{\today}
\maketitle

\begin{abstract}
We calculate the dynamic vortex mass for clean Fermi superfluids 
including
both $s$- and $d$-wave superconductors as a response to a vortex 
acceleration.
Assuming a finite quasiparticle mean free time, the vortex mass 
appears to 
be a tensor. The diagonal component dominates in the
limit of long mean free time while the off-diagonal mass takes over 
in the
moderately clean regime.
\end{abstract}

\pacs{PACS numbers: 74.20.-z, 74.60.Ge, 74.60.-w, 67.40.Vs, 67.57.Fg}
] \narrowtext

The vortex mass in superfluids and superconductors is a long standing
problem in vortex physics and remains to be an issue of controversies.  
There are different approaches to its definition.  One approach consists in  
calculating the vortex free energy increase due to the vortex velocity
\cite{Leggett}.  First used by Suhl \cite{Suhl} this approach 
yields the mass of the order of one quasiparticle mass (electron, in 
case of superconductor) per atomic layer (see Ref. \cite{Blatter} 
for an extensive review). Another approach is based on 
finding the force necessary to support an unsteady vortex motion. 
Identifying then the contribution to the force proportional to 
the vortex acceleration, one defines the vortex mass as a coefficient 
of proportionality.  This method was first applied for vortices in 
superclean superconductors in Ref. \cite{Kop/mass} and since then was 
used widely (see for example, \cite{KS,Simanek,Gaitonde}). The 
resulting  mass is of the order of the {\it total} mass of all 
particles within the area occupied by the vortex core. 
We will refer to this mass as to 
the {\it dynamic mass}.  The dynamic mass originates from the inertia 
of excitations localized in the vortex core and can also be calculated as 
the momentum carried by localized excitations \cite{Vol/mass}. It is much 
larger and thus it is much more important than the mass obtained from the 
energy considerations. 

In the present Letter we develop a regular microscopic approach for 
calculating the dynamic vortex mass in a general case of a finite 
relaxation time $\tau$ of nonequilibrium excitations produced by the 
moving vortex. 
Using the Boltzmann kinetic equation for quasiparticles localized in 
the vortex core, we derive the equation for the vortex dynamics which 
contains the inertia term together with all the forces acting on a 
moving vortex.
This approach is applied to both $s$- and $d$-wave superconductors. 
We find that dynamic mass displays a novel feature: it is a tensor whose
components depend on the quasiparticle relaxation time.
In $s$-wave superconductors, this tensor is diagonal in the superclean
limit. The diagonal 
mass decreases rapidly as a function of the mean free time, and the 
off-diagonal components dominate in the moderately clean regime. In 
$d$-wave superconductors, the transition from a diagonal mass (in the 
extremely clean, high-$\tau$ limit) to an off-diagonal mass tensor 
(in the moderately clean, low-$\tau$ regime) 
occurs via the intermediate universal regime of the flux flow predicted
in \cite{KopVol} (see also \cite{Makhlin,Kop/dwave}), where
both components are of the same order of magnitude. Our results agree with 
the previous work \cite{Kop/mass,KS,Vol/mass,Kop/dwave} in the limit 
$\tau \to \infty$.

{\it Forces on a moving vortex.}
The Boltzmann kinetic equation for
quasiparticles localized in the vortex core is
\begin{equation}
\frac{\partial f}{\partial t}+\frac{\partial f}{\partial \alpha}\frac{%
\partial \epsilon _{n}}{\partial \mu}-
\frac{\partial \epsilon _{n}}{\partial \alpha}\frac{%
\partial f}{\partial \mu}=\left( \frac{\partial f}{\partial t}\right)_{{\rm %
coll}}.  \label{Boltzmann/eq}
\end{equation}
Here $\epsilon _n(\alpha ,\mu )$ is the quasiclassical spectrum of a 
quasiparticle, which depends on canonically conjugated variables $\alpha $ and
$\mu$ where $\alpha $
is the angle of the particle velocity $\hat{{\bf v}}_\perp =\pm \hat {{\bf p}%
}_\perp$ in the plane perpendicular to the vortex axis, and $\mu =\mp
p_\perp b$ is the ``angular momentum'' defined through the impact 
parameter $b$ of a particle. The impact parameter is a good quantum 
number in the quasiclassical limit and so is $\mu$, even if there is 
no axial symmetry. The $\mp$ signs refer to particles and holes in the sense of
the normal-state Fermi surface with positive and negative effective masses
(in the plane perpendicular to the vortex axis), respectively. We 
take the coordinate $z$ axis along the vortex circulation $\hat {{\bf z}}={\rm 
sgn}\, (e){\bf B}/B$ where ${\bf B}$ is the magnetic induction. It can be 
shown  \cite{KopGeshBlatt} that, for the vortex dynamics, the 
Boltzmann-equation approach \cite{Stone} is equivalent to the full-scale 
Green function formalism \cite{KopKr,KL}. Eq. (\ref{Boltzmann/eq}) can
be derived for slow
in time variations of the distribution function such that the 
characteristic 
$\omega$ is small as compared to $T$ or to the bulk gap  $\Delta _0$.

For simplicity, we restrict ourselves to the case of well separated
vortices, 
$H\ll H_{c2}$, and consider the limit $\omega _c\tau \ll 1$ where 
$\omega _c=|e|H/m^*c$ is the cyclotron frequency. The limit 
$\omega _c\tau \ll 1$ is the most realistic
for superconductors; it also corresponds to uncharged superfluids. 
In this limit, the delocalized quasiparticles above the gap are in 
equilibrium with the heat bath and do not participate in the vortex 
motion \cite{KL}.

We seek the distribution function in the form 
$f=f^{(0)}(\epsilon _{n})+f_{1}$, where $f^{(0)}(\epsilon )=1-2n(\epsilon )$ 
and $n(\epsilon )$ is the Fermi function while 
$f_1$ describes the deviation from equilibrium. For a vortex moving with
the velocity ${\bf u}$ the angular momentum is time dependent: $\mu (t) 
=\left[({\bf r}-{\bf u}t)\times {\bf p}\right]\cdot \hat{{\bf z}}$. 
The driving term in Eq. (\ref{Boltzmann/eq}) is 
\begin{equation}
\frac{\partial f^{(0)}}{\partial t}=
\frac{\partial f^{(0)}}{\partial \epsilon}\frac{%
\partial \epsilon _{n}}{\partial \mu }\left( [{\bf p}_{\perp }\times{\bf u}]
\hat{{\bf z}}\right) \label{drive}.
\end{equation}

The spectrum of bound states in the vortex core has an anomalous 
chiral branch \cite{Caroli} $\epsilon _0 (\mu)$ whose energy varies
from $-\Delta _0$ to $\Delta _0$ crossing zero as the impact parameter of 
a particle in the core varies from $-\infty$ to $+\infty$. 
The anomalous branch corresponds to the
radial quantum number $n=0$. The energies of states with $n\ne 0$ are
separated from $\epsilon =0$ by a gap of the order of $\Delta_{0}$.  
Defining $\omega _n=\mp \partial \epsilon _n/\partial \mu$,
we ensure  $\omega _n$ to be positive on the branch $n=0$ for both 
particles with $\hat{{\bf v}}_\perp =\hat{{\bf p}}_\perp$ and holes with 
$\hat{{\bf v}} _\perp =-\hat{{\bf p}}_\perp$. Note that it is the Fermi 
velocity ${\bf v}_\perp$ rather than the momentum ${\bf p}_\perp $ 
that enters the quasiclassical equations which determine the excitation 
spectrum. Consequently, it is $\partial \epsilon _n/\partial b$ that has 
the same sign for particles and holes.

Multiplying Eq. (\ref{Boltzmann/eq}) by ${\bf p}_{\perp }/2$ and 
summming up over all the quantum numbers, we obtain
\begin{equation}
{\bf F}_{{\rm env}} ={\bf F}_{{\rm loc}} 
- \frac{\partial {\bf P}}{\partial t} -\pi \left( N_{e}-N_{h}\right) 
\tanh \left( \frac{\Delta _{0}}{2T}\right) [{\bf u}\times \hat{{\bf z}}] 
\label{Fenv1}
\end{equation}
where the l.h.s. of Eq. (\ref{Fenv1}) is the force from the environment on a 
moving vortex derived in \cite{KL}. It is the momentum transferred 
from excitations to the vortex:
\begin{eqnarray}
{\bf F}_{{\rm env}} =\frac{1}{2}\sum_{n}\int \frac{dp_{z}}{2\pi }\frac{%
d\alpha \,d\mu }{2\pi }[\hat{{\bf z}}\times {\bf p}_{\perp}]\frac{\partial
\epsilon _{n}}{\partial \mu}f_{1}.  \label{lhs}
\end{eqnarray}

The first term in the r.h.s. of Eq. (\ref{Fenv1}) is
the force exerted on the vortex by the heat bath via excitations
localized in the vortex core: 
\begin{equation}
{\bf F}_{{\rm loc}}=-\frac{1}{2}\sum_{n}\int \frac{dp_{z}}{2\pi}\frac{%
d\alpha \,d\mu }{2\pi }{\bf p}_{\perp } \left(\frac{\partial f}{\partial t}%
\right)_{{\rm coll}}.
\end{equation}
Its component parallel to ${\bf u}$ gives the friction force. The transverse 
component is nothing but the spectral flow force due to the localized
excitations \cite{KopVol/spflow}.
The second term in the r.h.s. of (\ref{Fenv1}) is the change in the vortex 
momentum defined as
\begin{equation}
{\bf P}=-\frac{1}{2}\sum _n\int \frac{dp_z}{2\pi}\frac{d\alpha \,d\mu}{2\pi} 
{\bf p}_\perp f_1 . \label{momentum}
\end{equation}

Finally, the third term in the r.h.s. of Eq. (\ref{Fenv1}) is obtained from
Eq. (\ref{drive}) (we assume here an $s$-wave pairing for simplicity).
The sum over $n$ contains only the
spectral branch with $n=0$ because $\partial \epsilon _n/\partial 
\mu$ is odd in $\mu$ for $n\ne 0$. It can be written as 
\begin{eqnarray}
&&-\pi N_{s}[{\bf u}\times \hat{{\bf z}}]-\pi N_{n}[{\bf u}\times \hat{%
{\bf z}}]  \nonumber \\
&&+\pi N\left[ 1-\tanh \left( \frac{\Delta _{0}}{2T}\right) \right] [{\bf u}%
\times \hat{{\bf z}}] . \label{term1}
\end{eqnarray}
Here we put $N=N_{e}-N_{h}$ and $N_{s}=N-N_{n}$ where $N_n$ is the 
normal density.
The term containing $N_{n}$ in Eq. (\ref{term1}) is the Iordanskii 
force\cite{Iordanskii} while the third one is a part of the spectral 
flow force due to
the excitations above the gap\cite{KopVol/spflow}. The last two terms 
in Eq. (\ref{term1}) together can be written as the transverse
force produced by the delocalized excitations scattered by the vortex
potential 
\begin{equation}
{\bf F}_{{\rm sc\,}\perp }=-\left[ {\bf u}\times {\bf \hat{z}}\right]
\int_{\left| \epsilon \right| >\Delta _{0}}\frac{d\epsilon }{4}\frac{%
\partial f^{\left( 0\right) }}{\partial \epsilon}\int \frac{%
dp_{z}}{2\pi }p_{\perp }^{3}\sigma _{\perp } \label{Fscat/perp}
\end{equation}
with the transverse cross section obtained in Ref. \cite{KopKr/scat} 
\begin{equation}
\sigma _{\perp }=\frac{\pi }{p_{\perp}}\left( 
\frac{\epsilon }{\sqrt{\epsilon ^{2}-\Delta _{0}^{2}}}-1\right) .
\label{crsec/sc/trans}
\end{equation}
Note that the similar longitudinal force can be neglected since it is smaller
than the transverse one by a factor $\left( p_{F}\xi \right) ^{-1}$  
\cite{KopKr/scat}. Eq. (\ref{Fenv1}) becomes 
\begin{equation}
{\bf F}_{{\rm env}} ={\bf F}_{{\rm loc}} -\frac{\partial {\bf P}}{\partial t}
-\pi N_{s}[{\bf u}\times \hat{{\bf z}}]+{\bf F}_{{\rm sc}}. 
\label{Fenv2}
\end{equation}

The equation of the vortex dynamics is obtained by the
variation of the superfluid free energy plus the energy 
in the external field with respect to the 
vortex  displacement. The variation of the
superfluid free energy gives the force from the environment, ${\bf 
F}_{{\rm env}}$, while the variation of the energy in the external field 
produces the external force. For charged superfluids the external 
force is just the Lorentz force \cite{LO} 
${\bf F}_L=(\Phi _0/c)[{\bf j}_s\times \hat{\bf z}]{\rm sgn}\, (e)$ 
where $\Phi _0 =\pi c/2|e|$ is the magnetic flux quantum.  This 
formula holds also for uncharged systems since the charge  
drops out giving the external force in the form 
${\bf F}_L=\pi N_s[{\bf v}_s\times \hat{\bf z}]$.
In the absence of pinning, ${\bf F}_L+{\bf F}_{{\rm env}}=0$ since 
the total energy is translationally invariant. 
Using Eq. (\ref{Fenv2}) the force balance takes the form
\begin{equation}
{\bf F}_{L}-\pi N_{s}[{\bf u}\times \hat{{\bf z}}]+{\bf F}_{{\rm loc}}
+{\bf F}_{{\rm sc}}=\frac{\partial {\bf P}}{\partial t}.
\label{forcebal1}
\end{equation}
In a Galilean invariant system, where $N$ is the total number of 
carriers, the $N_{s}$-term can be combined with the Lorentz force to give the 
Magnus force ${\bf F}_{M}=\pi N_{s}[\left( {\bf v}_{s}-{\bf u}\right) %
\times \hat{{\bf z}}]$. Eq. (\ref{forcebal1}) becomes
\begin{equation}
{\bf F}_{M}+{\bf F}_{{\rm loc}}+{\bf F}_{{\rm sc}}
=\frac{\partial {\bf P}}{\partial t}.  \label{forcebal2}
\end{equation}

The physical 
meaning of the Eq. (\ref{forcebal1}) [or (\ref{forcebal2})] is simple. 
The equation accounts for all the forces acting on a moving straight vortex 
line: the Magnus force from the superfluid component, the force due to the 
scattering of normal excitations ${\bf F}_{\rm sc}$, and the force from 
the heat bath transferred to the vortex by localized excitations, 
${\bf F}_{\rm loc}$. The r.h.s. of Eqs. (\ref{forcebal1}, \ref{forcebal2}) 
comes from the inertia of excitations
localized in the vortex core and is identified as the change in the
vortex momentum.  The definition of Eq. (\ref{momentum}) is similar 
to that used in  Refs. \cite{Vol/mass,Stone}. 
Note that the delocalized quasiparticles do not 
contribute to the vortex momentum in the limit $\omega _c\tau \ll 1$ because 
their distribution is equilibrium \cite{KL} with $f_1=0$.

{\it Vortex mass.}
Having defined the vortex momentum, we calculate the 
vortex mass. The distribution function of localized particles is\cite{KL} 
\begin{equation}
f_{1}=-\frac{df^{(0)}}{d\epsilon }\left[\gamma _{{\rm H}}\left( {\bf 
u}\cdot {\bf p}_{\perp }\right) \pm \gamma _{{\rm O}}\left( [{\bf u}\times 
{\bf p}_{\perp }]\cdot \hat{{\bf z}}\right) \right] \label{distrib}
\end{equation}
where $\gamma _{\rm H}$ and $\gamma _{\rm O}$ are to be found from 
Eq. (\ref{Boltzmann/eq}). The vortex momentum becomes 
$P_i =M_{ik} u_{k}$; it has both longitudinal and transverse components 
with respect to the vortex velocity.

For a vortex with the symmetry not less than the four-fold, the 
effective
mass tensor per unit length is $M_{ik}=M_\parallel \delta _{ik}- 
M_\perp e_{ikj}{\hat z}_j$ where $M_\parallel =M_{\parallel \, e}
+M_{\parallel\, h}$ and $M_\perp =M_{\perp \, e}-M_{\perp \, h}$ with
\begin{equation}
M_{\parallel\, e,h}=\frac{1}{4}\sum_{n}\int _{e,h}\frac{dp_{z}}{2\pi} 
\frac{d\mu d\alpha}{2\pi}\frac{df^{(0)}}{d\epsilon }\,
p_{\perp }^{2}\gamma _{{\rm H}} \label{mass/tens}
\end{equation}
and the same expression for $M_{\perp\, e,h}$ with $\gamma _{\rm H}$ 
replaced with $\gamma _{\rm O}$.
The $e,h$ subscripts indicate the integrations over the
electron and hole parts of the Fermi surface, respectively. Only the
branch with $n=0$ gives the contribution to the transverse 
mass because $\gamma _{\rm O}$ is odd in $\mu$ for $n\ne 0$.

Eq. (\ref{momentum}) does not contain any explicit contribution from 
the so called ``backflow mass''. The backflow vortex mass was introduced 
\cite{Baym} to account for an apparent similarity between a vortex core and an 
impenetrable cylinder immersed into a fluid. A hydrodynamic dipole 
counterflow would arise around the
cylinder moving with respect to the fluid giving rise to the associated
backflow mass\cite{Landau}. We see, however, that the total inertial 
force comes from the
change in the momentum of localized excitations only. In a sense,
Eq. (\ref{momentum}) takes care of all sources of
the vortex mass including the possible backflow effect.

We also comment on arguments of Ref. \cite{Gaitonde} that the 
Coulomb interactions from charge 
variations produced by the moving vortex would screen the dynamic mass 
in charged superfluids. It is well known that moving vortex induces a
non-equilibrium chemical potential which compensates the changes in the
charge density (see, for example, \cite{GK,LO}) .  Therefore, the vortex 
motion does not violate the charge neutrality and the abovementioned 
screening effects are absent. (Note also that such a screening would  first 
of all have screened out the whole vortex response to an electric 
current.)

{\it $s$-wave superconductors.}
For $s$ wave superconductors, $\omega _n$ is independent of $\alpha$.
If the collision integral is taken in the $\tau$-approximation
$\left(\partial f/\partial t\right)_{\rm coll}=-f_1/\tau _n$, 
the solution to Eq. (\ref{Boltzmann/eq}) for a steady vortex motion is 
\cite{KL} 
\begin{equation}
\gamma _{{\rm H}}=\frac{\omega _{n}^{2}\tau _{n}^{2}}{\omega _{n}^{2}\tau%
_{n}^{2}+1};~
\gamma _{{\rm O}}=\frac{\omega _{n}\tau _{n}}{\omega _{n}^{2}\tau _{n}^{2}+1} 
\label{gamma/swave}
\end{equation}
where $\omega _n \sim T_c^2/E_F$. At low temperatures one has
\[ M_{\parallel\, e,h}=\pi N_{e,h} \left\langle \frac{\gamma _{\rm H}}
{\omega _0}\right\rangle ,\;
M_{\perp\, e,h}=\pi N_{e,h} \left\langle \frac{\gamma _{\rm O}}
{\omega _0}\right\rangle \]
where $\left\langle \ldots \right\rangle =(\pi /V_F)\int p_\perp ^2\, 
dp_z (\ldots )$ is the averaging over the Fermi surface with the volume 
$V_F$. The mass is of the order of the mass of electrons in the area 
occupied by the vortex core $M\sim \pi \xi _{0}^{2}mN$.
In the superclean limit $T_c^2\tau /E_F \gg 1$ where
$\omega _n\tau \gg 1 $ the mass tensor is diagonal $%
M_{ik}=M_\parallel \delta _{ik}$, and we obtain the known
result \cite{Kop/mass,KS,Vol/mass}: 
$M_{\parallel\, e,h} =
\pi N_{e,h} \left\langle \omega _0^{-1}\right\rangle $.

The mass decreases with $\tau $. In the moderately clean regime 
$T_c^2\tau /E_F \ll 1$ where 
$\omega _n\tau \ll 1$, the diagonal component vanishes as $\tau
^{2}$, and the mass tensor is dominated by the off-diagonal part. If 
$\tau _0$
is independent of energy
\[
M_{\perp\, e,h}=\pi \tanh \left(\frac{\Delta _{0}}{2T}\right) N_{e,h}
\left\langle \tau _{0}\right \rangle .
\]

The vortex mass is not a constant quantity for a given system: it may
depend  on the frequency $\omega $ of the external drive.  Indeed,
for $\omega $ comparable with $\omega _{0}$ and $\tau ^{-1}$, both
the forces and the effective mass 
acquire a frequency dispersion. From Eq. (\ref{Boltzmann/eq}) one finds
\[
\gamma _{{\rm H}}=\frac{\omega _{n}^{2}\tau _{n}^{2}}{\omega _{n}^{2}\tau
_{n}^{2}+(1-i\omega \tau _n)^2};~ \gamma _{{\rm O}}=\frac{\omega _{n}\tau
_{n}(1-i\omega \tau _n)}{\omega _{n}^{2}\tau _{n}^{2}+(1-i\omega \tau _n)^2}.
\]

{\it $d$-wave superconductors.}
We take the $d$-wave order parameter in the 
form $\Delta =\Delta _0(\rho ,\phi )\sin (2\alpha )e^{i\phi}$. The 
interlevel spacing $\omega _n(\alpha )$ for the quasiclassical 
spectrum $\epsilon _n(\alpha , \mu )$ of particles in the vortex core
depends now on the momentum direction $\alpha$.  

One distinguishes three different regimes for systems with different purity. 
In the order of increasing $\tau$ we have (i) the moderately clean 
regime, (ii)  the universal regime, and (iii) the extremely
clean limit.  The moderately clean 
limit corresponds to $T_c^2\tau /E_F\ll 1$. In this case, the factors 
$\gamma$ in Eq. (\ref{distrib}) are
$\gamma _{\rm H}=\omega _n^2(\alpha )\tau _n^2$ and
$\gamma _{\rm O}=\omega _n(\alpha )\tau _n$. The mass tensor has 
qualitatively the same behavior as for an $s$-wave superconductor.

In the superclean limit $T_c^2\tau /E_F\gg 1$, the specifics of a $d$-wave 
superconductor is more pronounced \cite{KopVol,Kop/dwave}.
Consider temperatures $T\ll T_{c}\sqrt{H/H_{c2}}$. At such
temperatures, only localized states on the anomalous branch are 
excited. We need the quasiclassical
spectrum for momentum directions near the gap nodes. The distance 
between the quasiclassical levels on the anomalous branch near the 
gap nodes is \cite{Kop/dwave}
\begin{equation}
\omega _0(\alpha )=8\Omega _0\alpha ^2+\omega _c/2
\end{equation}
where $\Omega _0\sim T_c^2/E_F$,
and $\omega _c=|e|Hv_\perp /cp_\perp$ is the cyclotron frequency. For 
comparison, $\omega _c/\Omega _0 \sim H/H_{c2}$.
The factors $\gamma _{\rm H}$ and $\gamma _{\rm O}$ in Eq. (\ref{distrib}) 
satisfy the set of equations
which follow from Eq. (\ref{Boltzmann/eq}) \cite{KopVol,Kop/dwave}
\begin{eqnarray}
\frac{\partial \gamma _{\rm O}}{\partial \alpha }-\gamma _{\rm H}
-U(\alpha )\gamma _{\rm O}+1 &=&0,  \nonumber \\
\frac{\partial \gamma _{\rm H}}{\partial \alpha }+\gamma _{\rm O}
-U(\alpha )\gamma _{\rm H} &=&0.  \label{eq/gamma}
\end{eqnarray}
Here $U(\alpha )=\left[ \omega _0(\alpha )\tau _0\right]^{-1}$.
The masses are 
\begin{eqnarray*}
M_{\parallel ,\, \perp} &=&\int \frac{dp_{z}}{2\pi }\,p_{\perp }^{2}
\int_{-\pi /4}^{\pi /4}\frac{%
d\alpha }{2\pi }\int \omega _{0}\left( \alpha \right) d\mu \frac{df^{(0)}}{%
d\epsilon }\frac{\gamma _{{\rm H},\, {\rm O}} \left( \alpha \right) }%
{\omega _{0}\left( \alpha \right) } \\
&=&\frac{\tau}{\pi}\int \frac{dp_{z}}{2\pi }\,p_{\perp }^{2}
\int_{-\pi /4}^{\pi /4} d\alpha \frac{dF}{d\alpha}
\gamma _{{\rm H},\, {\rm O}}
\end{eqnarray*}
where $F\left( \alpha \right) =\int_{0}^{\alpha }U(\alpha ^\prime )\, 
d\alpha ^\prime$ and $\tau \equiv \tau _0(\alpha =0)$.
The integral over $d\alpha $ is determined by the angles $\alpha \sim 
\sqrt{H/H_{c2}}$. For this range, the terms with $U(\alpha )$ in 
Eq. (\ref{eq/gamma}) dominate and the factors 
$\gamma _{{\rm H},\, {\rm O}}$ are \cite{KopVol,Kop/dwave}
\begin{equation}
\gamma _{{\rm H}}=\frac{\cosh \lambda \;e^{F\left( \alpha \right)}}%
{2\sinh ^{2}\lambda +1}, \;
\gamma _{{\rm O}}=\frac{\sinh \lambda \; e^{F\left( \alpha \right)}}%
{2\sinh ^{2}\lambda +1}.
\label{gammas}
\end{equation}
Here
\begin{equation}
\lambda =\int_{0}^{\infty }\frac{d\alpha ^{\prime }}{\omega _{0}\left(
\alpha ^{\prime }\right) \tau }=\frac{\pi }{4E_{0}\tau }\label{lambda}
\end{equation}
and $E_{0}=\sqrt{\Omega _{0}\omega _{c}}$
is the distance between the exact quantum levels in the vortex core. 
They can be found by semi-classical quantization of the azimuthal motion of the
particle having the spectrum $\epsilon _0(\alpha ,\mu )$ \cite{Kop/dwave}.
Finally, the masses become
\begin{eqnarray*}
M_{\parallel\, e,h } &=&2N_{e,h}\left\langle
\frac{2\tau \tanh \lambda }{\tanh ^{2}\lambda +1}\right\rangle , \\
M_{\perp \, e,h} &=&2N_{e,h}\left \langle
\frac{2\tau \tanh ^{2}\lambda }{\tanh ^{2}\lambda +1} \right\rangle .
\end{eqnarray*}

The universal regime \cite{KopVol} corresponds to $\lambda \gg 1$, i.e.,  
$E_0\tau \ll 1$.
In this limit, the spacing between the exact quantum levels is much 
smaller than the relaxation rate, while the average distance between the 
quasiclassical levels is larger than the relaxation rate, 
$\Omega _0 \tau \gg 1$. Here, the both masses are of the same
order of magnitude $M_{\parallel }\sim M_{\perp }\sim N\tau $.
In the extremely clean limit when $\lambda \ll 1$, i.e., $E_0\tau \gg 1$, 
one obtains
$M_{\parallel\, e,h}=\pi N_{e,h}\left\langle E_{0}^{-1}\right\rangle $
which agrees with the result of Ref. \cite{Vol/mass,Kop/dwave}. The 
transverse mass is small $M_{\perp }\sim \lambda M_{\parallel }$.

In conclusion, we have calculated the dynamic vortex-mass tensor as a 
response to a slow vortex acceleration. 
The diagonal component dominates in the limit of a very long mean 
free time. In the opposite limit of fast relaxation, the off-diagonal mass is 
more important. The order
of magnitude of the dynamic mass is the mass of all the quasiparticles in the 
area occupied by the vortex core. This mass is much larger than both the 
electrodynamic and the Suhl masses and is associated with the inertia of 
excitations localized in the vortex core.

We are grateful to G. Blatter, V. Geshkenbein, E. Sonin, and G. Volovik 
for stimulating discussions. This work was supported by the
U.S. Department of Energy, BES-Materials Sciences,
under contract \# W-31-109-ENG-38 and by 
the NSF-Office of the
Science and Technology  Center under contract No. DMR91-20000.
NK also acknowledges support from the Russian Foundation for
Fundamental Research grant No. 96-02-16072 and from the program 
``Statistical Physics'' of the Ministry of Science of Russia 
and from the Swiss National Foundation cooperation grant 7SUP J048531.


\begin{references}
\bibitem{Leggett} J.-M. Duan and A. J. Leggett, Phys. Rev. Lett. {\bf 
68}, 1216 (1992).

\bibitem{Suhl}  H. Suhl, Phys. Rev. Lett. {\bf 14}, 226 (1965).

\bibitem{Blatter}  E. B. Sonin, V. B. Geshkenbein, A. van Otterlo, 
and G. Blatter, Phys. Rev. B {\bf 57}, 575 (1998).

\bibitem{Kop/mass}  N. B. Kopnin, Pis'ma Zh. Eksp. Teor. Fiz. {\bf 27}, 417
(1978) [JETP Lett. {\bf 27}, 390 (1978)].

\bibitem{KS}  N. B. Kopnin and M. M. Salomaa, Phys. Rev. B, {\bf 44}, 
9667 (1991).

\bibitem{Simanek}  E. \v {S}im\'{a}nek, Journ. Low Temp. Phys. {\bf 100}, 
1 (1995).

\bibitem{Gaitonde}  D. M. Gaitonde and T. V. Ramakrishnan, Phys. Rev. 
B {\bf 56}, 11 951 (1997).

\bibitem{Vol/mass}  G. E. Volovik, Pis'ma Zh. Eksp. Teor. Fiz. {\bf 
65}, 201 (1997) [JETP Lett. {\bf 65}, 217 (1997)].

\bibitem{KopVol}  N. B. Kopnin and G. E. Volovik, Phys. Rev. Lett. 
{\bf 79}, 1377 (1997).

\bibitem{Makhlin} Yu. G. Makhlin, Phys. Rev. B, {\bf 56},11 872 
(1997). 

\bibitem{Kop/dwave}  N. B. Kopnin, Phys. Rev. B {\bf 57}, 11 775 
(1998).

\bibitem{KopGeshBlatt}  N. B. Kopnin, V. B. Geshkenbein, and G. 
Blatter, to be published.

\bibitem{Stone}  M. Stone, Phys. Rev. B {\bf 54}, 13 222 (1996).

\bibitem{KopKr}  N. B. Kopnin and V. E. Kravtsov, Pis'ma Zh. Eksp. 
Teor. Fiz. {\bf 23}, 631 (1976) [JETP Lett. {\bf 23}, 578 (1976)].

\bibitem{KL}  N. B. Kopnin and A. V. Lopatin, Phys. Rev. B {\bf 51}, 
15 291 (1995).

\bibitem{Caroli}  C. Caroli, P. G. de Gennes, and J. Matricon, Phys. 
Lett. {\bf 9}, 307 (1964).

\bibitem{KopVol/spflow}  N. B. Kopnin, G. E. Volovik, and \"{U}. 
Parts, Europhys. Lett. {\bf 32}, 651 (1995).

\bibitem{Iordanskii}  S. V. Iordanskii, Ann. Phys.{\bf \ 29}, 335 
(1964).

\bibitem{KopKr/scat}  N. B. Kopnin and V. E. Kravtsov, Zh. Eksp. 
Teor. Fiz. {\bf 71}, 1644 (1976) [Sov. Phys. JETP {\bf 44}, 861 (1976)].

\bibitem{Baym}  G. Baym and E. Chandler, Journ. Low Temp. Phys. {\bf 50}, 57
(1983).

\bibitem{Landau} L. D. Landau and E. M. Lifshits, {\it Fluid mechanics}
(Pergamon Press, N.Y. 1987).

\bibitem{GK}  L. P. Gorkov and N. B. Kopnin, Usp. Fiz. Nauk {\bf 
116}, 413 (1975) [Sov. Phys. Uspekhi {\bf 18}, 496 (1976)].

\bibitem{LO}  A. I. Larkin and Yu. N. Ovchinnikov, in {\it Nonequilibrium
Superconductivity}, edited by D. N. Langenberg and A. I. Larkin 
(Elsevier Science Publishers, 1986), p. 493.

\end{references}
\end{document}